\documentclass[12pt]{iopart}
\usepackage{graphicx}
\usepackage{iopams}

\begin{document}

\title{Tables of experimental branching ratios of Auger-type post collisional ionization of rare gases}
\author{C C Montanari and J E Miraglia}

\address{Instituto de Astronom\'{\i}a y F\'{\i}sica del Espacio,
Consejo Nacional de Investigaciones Cient\'{\i}ficas y T\'ecnicas
and Universidad de Buenos Aires, C1428EGA, Buenos Aires, Argentina}
\address{Facultad de Ciencias Exactas y Naturales, Universidad de Buenos Aires, Buenos Aires, Argentina}
\ead{mclaudia@iafe.uba.ar}

\begin{abstract}
When an electron is emitted from a sub-valence shell, a vacancy is
created and there is a not-null probability for different post
collisional ionization (PCI) processes giving rise to a higher
target charge state. It is reasonable to consider that PCI is a
time-delayed electron emission and, therefore, independent of the
projectile. We include the tables of the branching ratios of $0-6$
post-collisionally emitted electrons after single ionization of Ne,
Ar, Kr and Xe in different inner shells. These values have already
been employed with good results in previous calculations of multiple
ionization of rare gases by electrons, positron, protons,
antiprotons and different positive ions (see the text and references
therein). These tables and explanation is included in [Montanari C C
and Miraglia J E 2014, \textit{J. Phys. B: At. Mol. Opt. Phys.} {\bf
45}, 105203].
\end{abstract}

\section{Introduction}

The experimental and theoretical work on post-collisional electron
emission provides detailed distributions of the different channels
that contribute to each final charge state. The inclusion of PCI
within multiple ionization (MI) calculations needs less detailed
information, just the branching ratio of charge state distribution
$F_{j,k}$, with $j$ being the orbital of the initial single vacancy
and $k$ the number of electrons emitted in PCI.

Following \cite{Spranger04,Cavalcanti02,MM1,AP,galassi},
post-collisional MI is included using the accurate experimental data
of single-photoionization ending in multiple-charged ions. While
collecting these experimental values two points must be ensured: the
sub-shell of the initial vacancy and that it is a single ionization
followed by PCI (not double or triple photoionization, for example).
In this way final target charge states can be attributed to
post-collisional events.

As in our previous works \cite{MM1,AP}, we use this unitarity of the
branching ratios (i.e. $1=\sum_{k=0}^{k_{max}}F_{j,k}$) to introduce
them in multinomial expression as
\begin{equation}
P_{(q_{j})}(b)=\left(\matrix {N_{j} \cr q_{j}} \right) \
\left[p_{j}(b)\ \sum_{i=0}^{i_{max}}F_{j,i}
\right]^{q_{j}}[1-p_{j}(b)]^{N_{j}-q_{j}}. \label{binomial_PCI}
\end{equation}

The MI probabilities including direct and PCI are obtained from
(\ref{binomial_PCI}), expanding the different terms and putting
together those that end in the same number of final emitted
electrons (details in section 2.3 of \cite{MM1}). This rearrangement
gives rise to $\mathsf{P}_{(n)}^{PCI}$, the probabilities of exactly
$n$ emitted electrons, part of them in direct ionization and part in
PCI.  This formalism has already been employed with good results in
previous calculations of multiple ionization including direct and
post-collsional contributions
\cite{MM1,AP,icpeacproc11,B2,sigaud13,archubi,erare,positron}.

\section{Branching ratios} \label{branching}

\begin{table}[tbp]
\caption{Branching ratios for multiple ionization of Ne and Ar.
$F_{j ,i}$ is the yield of single photoionization of the $j
$-subshell followed by the post-collisional emission of $i$
electrons from outer shells, with the final charge state being
$i+1$.} \label{tableNeAr}
\begin{indented}
\lineup
\item[]
\begin{tabular}{@{}lcccccccccc}
\br
           &{\bf Ne}     &           &             &  & & {\bf Ar}    &             &                  &                   &         \\
\mr
           &{\bf 1s$^a$}&{\bf 2s$^b$}& {\bf 2p$^b$}&  & &{\bf 1s$^c$} &{\bf 2s$^d$} &{\bf 2p$^{ d}$}   & {\bf 3s$^{c,d,e}$}  &{\bf 3p} \\
$F_{j ,0}$ & 0.0193     & 0.98       & 0.98        &  & & 0.070       &  0.000      & 0.005            & 1.00              & 1.00    \\
$F_{j ,1}$ & 0.921      & 0.02       & 0.02        &  & & 0.105       &  0.010      & 0.863            & 0.00              & 0.00    \\
$F_{j ,2}$ & 0.0571     & 0.00       & 0.00        &  & & 0.078       &  0.890      & 0.128            & 0.00              & 0.00    \\
$F_{j ,3}$ & 0.0028     & 0.00       & 0.00        &  & & 0.427       &  0.100      & 0.003            & 0.00              & 0.00    \\
$F_{j ,4}$ & 0.000      & 0.00       & 0.00        &  & & 0.250       &  0.000      & 0.001            & 0.00              & 0.00    \\
$F_{j ,5}$ & 0.000      & 0.00       & 0.00        &  & & 0.103       &  0.000      & 0.000            & 0.00              & 0.00    \\
$F_{j ,6}$ & 0.000      & 0.00       & 0.00        &  & & 0.024       &  0.000      & 0.000            & 0.00              & 0.00    \\
\br
\end{tabular}
\end{indented}

\begin{indented}
\item[]
\textbf{For Ne:} %
{\bf 1s:}$^{\rm a}$, Landers \etal \protect\cite{landers}; similar
experimental data (differences within 2$\%$) can be found in Morgan
\etal\protect\cite{morgan}, Saito \etal \protect\cite{saito} and
Carlson \etal \protect\cite{carlson65Ne}
(the latter corrected to exclude L-shell ionization); %
{\bf 2s} and {\bf 2p:}$^{\rm b}$, theoretical shake-off contribution
\cite{Mukoyama,CN1}. $F_{2s ,1}\approxeq 0.03-0.06$ has been
suggested by Schmidt \etal \cite{schmidt}.
\item[] \textbf{For Ar:} %
{\bf 1s:} $^{\rm c}$ Carlson \etal \cite{carlson}; {\bf 2s} and {\bf
2p:} $^{\rm d}$ Br\"{u}nken \etal \protect\cite{brunken}; similar
values for 2p-photoionization have been reported by Viefhaus \etal
\protect\cite{viefhaus}; {\bf 3s:} different works, $^{\rm c}$
Carlson \etal \cite{carlson}, $^{\rm d}$ Br\"{u}nken \etal
\protect\cite{brunken} and recently $^{\rm e}$ Karamatskos \etal
\cite{kara}.
\end{indented}
\end{table}
Carlson, Krause and co-workers have been pioneers in the
experimental and theoretical studies on multiple photoionization
\cite{carlson,krause,carlson65Ne,CN1,CN2}. They use X-ray tubes to
get photons in certain energy range close to that of the initial
vacancies, so these measurements include outer-shell ionization too.
The advent of the new experimental techniques in photoionization
research brought a breakthrough in this field. Coincident
measurements allow to accurately determine the total yield of
Auger-electron emission processes for vacancies in different shells
of various atoms and to separate multiple-photoionization from
single-photoionization followed by PCI (see the compilation of
literature in tables \ref{tableNeAr}, \ref{tableKr}  and
\ref{tableXe}). However, it is fair to note that the earlier
measurements by Carlson and collaborators in the '60s, corrected to
exclude outer-shell ionization, are in good agreement with recent
ones using COLTRIMS techniques (see for example the corrected values
by Krause \etal \cite{krause} for Kr, which are quite close to
recent values by Tamenori \etal \cite{tamenori04} and Armen \etal
\cite{armen}).

In tables \ref{tableNeAr}, \ref{tableKr} and \ref{tableXe} we
display the branching ratios used in this work for PCI of 0-6
electrons and all the shells of Ne, Ar, Kr and Xe. These values have
already been employed with good results in previous calculations of
multiple ionization of rare gases by protons, antiprotons and
different positive ions
\cite{MM1,AP,icpeacproc11,B2,sigaud13,archubi,erare,positron}. A
detailed discussion on the experimental branching ratios and the
comparison of the different data available in the literature is
included in \cite{MM1}. This review has been updated and completed
in the present contribution. For example in tables \ref{tableKr} and
\ref{tableXe} we display the data for the deepest shells of Kr and
Xe. The importance of these contributions in MI is analyzed in
Tavares \etal \cite{sigaud13}.

\begin{table}[tbp]
\caption{Branching ratios for multiple ionization of Kr. Notation as
in table \ref{tableNeAr}.} \label{tableKr}
\begin{indented}
\lineup
\item[]
\begin{tabular}{@{}lcccccccc}
\br
{\bf Kr}   &               &               &               &               &               &               &                &         \\
\mr
           &{\bf 1s$^f$}   &{\bf 2s$^{f}$} & {\bf 2p$^{g}$}&{\bf 3s$^{h}$} &{\bf 3p$^{h}$} &{\bf 3d$^{h}$} & {\bf 4s$^{h}$} &{\bf 4p} \\
$F_{j ,0}$ &$<$0.01         &  0.00         &  0.00         &  0.00         &  0.00         & $<$0.01       & 1.00           & 1.00    \\
$F_{j ,1}$ &   0.01         &  0.00         &  0.01         &  $<$0.02      &  $<$0.02      & 0.670         & 0.00           & 0.00    \\
$F_{j ,2}$ &   0.04         &  $<$0.01      &  0.03         &  0.12         &  0.623        & 0.320         & 0.00           & 0.00    \\
$F_{j ,3}$ &   0.19         &  0.02         &  0.19         &  0.66         &  0.333        & $<$0.01       & 0.00           & 0.00    \\
$F_{j ,4}$ &   0.21         &  0.156        &  0.35         &  0.21         &  $<$0.02      & $<$0.01       & 0.00           & 0.00    \\
$F_{j ,5}$ &   0.17         &  0.30         &  0.31         &  0.00         &  0.00         & 0.00          & 0.00           & 0.00    \\
$F_{j ,6}$ &   0.13         &  0.35         &  0.11         &  0.00         &  0.00         & 0.00          & 0.00           & 0.00    \\
\br
\end{tabular}
\end{indented}

\begin{indented}
\item[] \textbf{For Kr:} %
{\bf 1s}$^{f}$ El-Shermi \etal \protect\cite{elshermi}
{\bf 2s}$^{f}$ El-Shermi \etal \protect\cite{elshermi}; %
{\bf 2p}$^{g}$ Morishita \etal \protect\cite{morishita}, other
values can be found in Armen \etal \protect\cite{armen} and in
El-Shermi \etal \protect\cite{elshermi}; %
{\bf 3s}$^{h}$ Tamenori \etal \protect\cite{tamenori04}; %
{\bf 3p}$^{h}$ Tamenori \etal \protect\cite{tamenori04}, the values
displayed are weighted mean values of $3p_{3/2}$ and $3p_{1/2}$;
alternative values (differences less than $10\%$) can be found in
Br\"{u}nken \etal \protect\cite{brunken}, Matsui \etal \cite{matsui}
and Armen \etal \cite{armen}; %
{\bf 3d}$^{h}$ Tamenori \etal \protect\cite{tamenori04}, other
similar data in Br\"{u}nken \etal \protect\cite{brunken} and Saito
\etal \cite{saito97}, recent theoretical values by Zeng \etal \cite{zeng} are in good agreement too; %
{\bf 4s, 4p}$^{h}$ Tamenori \etal \protect\cite{tamenori04} show
that the energy threshold of Kr$^{2+}$ is above Kr$^{+}(4s^{-1})$,
so the decay from Kr$^{+}(4s^{-1})$ to Kr$^{2+}$ is not
energetically possible .
\end{indented}
\end{table}

\begin{table}[tbp]
\caption{Similar to \ref{tableNeAr} and \ref{tableKr} for Xe
target.} \label{tableXe}
\begin{indented}
\lineup
\item[]
\begin{tabular}{@{}lccccccccccc}
\br
{\bf Xe}   &               &               &               &              &                &               &               &               &              &           &         \\
\mr
           &{\bf 1s$^f$}   &{\bf 2s$^{f}$} & {\bf 2p$^{f}$}&{\bf3s}$^{k}$ &{\bf 3p}$^{l}$  &{\bf 3d}$^{l}$ &{\bf 4s}$^{k}$ &{\bf 4p}$^{m}$ &{\bf 4d}$^{p}$ & {\bf 5s} &{\bf 5p} \\
$F_{j ,0}$ & $<$0.01       &  $<$0.01      &  $<$0.01      &  0.00        &  0.00          &  0.00         &  0.00         & 0.00          & 0.00          & 1.00      & 1.00    \\
$F_{j ,1}$ &   0.019       &  $<$0.01      &  $<$0.01      &  0.00        &  0.00          &  0.00         &  0.01         & 0.05          & 0.80          & 0.00      & 0.00    \\
$F_{j ,2}$ &   0.011       &  $<$0.01      &  $<$0.01      &  0.003       &  0.00          &  0.049        &  0.165        & 0.89          & 0.20          & 0.00      & 0.00    \\
$F_{j ,3}$ &   0.043       &  0.012        &  0.042        &  0.007       &  0.00          &  0.43         &  0.774        & 0.06          & 0.00          & 0.00      & 0.00    \\
$F_{j ,4}$ &   0.058       &  0.035        &  0.035        &  0.024       &  0.13          &  0.32         &  0.051        & 0.00          & 0.00          & 0.00      & 0.00    \\
$F_{j ,5}$ &   0.114       &  0.068        &  0.063        &  0.126       &  0.27          &  0.17         &  0.00         & 0.00          & 0.00          & 0.00      & 0.00    \\
$F_{j ,6}$ &   0.143       &  0.104        &  0.097        &  0.426       &  0.43          &  0.028        &  0.00         & 0.00          & 0.00          & 0.00      & 0.00    \\
\br
\end{tabular}
\end{indented}

\begin{indented}
\item[] \textbf{For Xe:} %
{\bf 1s}$^{f}$ El-Shermi \etal \protect\cite{elshermi}; %
{\bf 2s}$^{f}$ El-Shermi \etal \protect\cite{elshermi}; %
{\bf 2p}$^{f}$ El-Shermi \etal \protect\cite{elshermi}; %
{\bf 3s} and {\bf 4s}$^{k}$ Kochur \etal  \protect\cite{kochur94},
theoretical calculation; no experimental data available; %
{\bf 3p}$^{l}$ Saito and Suzuki \protect\cite{saito92}; %
{\bf 3d}$^{l}$ Saito and Suzuki \protect\cite{saito92}; similar
experimental values have been reported in \protect\cite{tamenori02};
also in agreement with differential data for $3d_{5/2}$ initial vacancy in
\protect\cite{matsui,tamenori02,partanen05}.%
{\bf 4p}$^{m}$ Hikosaka \etal \protect\cite{hikosaka2}; the
theoretical results by
Kochur \etal  \protect\cite{kochur94} are in good agreement with these experimental values; %
{\bf 4d}$^{p}$ Hayaishi \etal \protect\cite{haya90} and K\"{a}mmerling \etal \protect\cite{kammerling};
the Xe N-shell values by Carlson \etal \protect\cite{carlson} include O-shell ionization and are not presented in this table. %
\end{indented}
\end{table}

The selection criteria for the branching ratios takes into account
the latest experimental techniques, coincident measurements, the
comparison between experimental and theoretical values, and the
convergence of the data around a certain value. Only in two cases,
$3s$ and $4s$ initial vacancy of Xe, no experimental values were
found so we include in table \ref{tableXe} the theoretical results
by Kochur \etal \cite{kochur94}. In this respect, it should be
mentioned that the calculated branching ratios for Xe by Kochur
\etal \cite{kochur94} for the other sub-shells are in good agreement
with the experimental ones displayed in table \ref{tableXe}.

A point of discussion is the possibility of PCI following the
initial ionization of a valence electron \cite{AP}. Auger emission
after ionization of the $ns$-subshell (with $n$ being the quantum
number of the valence shell) is not energetically possible, even
when there are 6 outer electrons in the $np$-subshell \cite{kochur}.
Very recently Karamatskos \etal \cite{kara} underlined this by
saying that Ar with a $3s$ vacancy is a no autoionizing state (only
spontaneous radiative decay is possible). This is expressed as
$F_{n,i}=\delta_{i,0}$ (no electron is emitted in PCI) in the last
two columns of tables \ref{tableNeAr}, \ref{tableKr} and
\ref{tableXe}. However the discussion of other possible mechanisms
(non-Auger) of second electron emission in valence shell ionization
is open, for example shake-off processes due to the change in the
target potential \cite{CN1,CN2,Mukoyama,Kochur06}, or interatomic
coulomb decay in dimers \cite{janhke,marburger,kim,yan}.

The case of Ne is different from the others in many aspects. In
table \ref{tableNeAr} we display the branching ratios after L-shell
ionization (Ne $2s$ and $2p$). These values are based on shake-off
calculations by Mukoyama \etal \cite{Mukoyama}, as explained in
\cite{AP}. We found that above 1 keV the double ionization cross
section of Ne cannot be explained by direct double ionization or by
K-shell ionization followed by PCI \cite{AP,MM1} of one more
electrons. This may be an indirect evidence of PCI following
valence-shell ionization for this atom. This subject remains open to
further research.

\ack{This work was partially supported by the following Argentinean
institutions: Consejo Nacional de Investigaciones Cient\'{\i}ficas y
T\'{e}cnicas, Agencia Nacional de Promoci\'{o}n Cient\'{\i}fica y
Tecnol\'{o}gica, and Universidad de Buenos Aires.}

\end{document}